\documentclass[twocolumn,english,journal]{IEEEtran}
\ifCLASSINFOpdf
\else
\fi

\usepackage[T1]{fontenc}
\usepackage{babel}
\usepackage{array}
\usepackage{calc}
\usepackage{amsmath}
\usepackage{amsthm}
\usepackage{amssymb}
\usepackage{graphicx}
\usepackage{cite}
\usepackage{algorithm}
\usepackage{subfigure}
\usepackage{color}

\usepackage{enumitem}
\usepackage{multirow}
\usepackage{hhline}
\usepackage{float}
\usepackage[justification=centering]{caption}

\usepackage[unicode=true,
bookmarks=true,bookmarksnumbered=true,bookmarksopen=true,bookmarksopenlevel=1,
breaklinks=false,pdfborder={0 0 0},backref=false,colorlinks=false]
{hyperref}
\hypersetup{pdftitle={Your Title},
	pdfauthor={Your Name},
	pdfpagelayout=OneColumn, pdfnewwindow=true, pdfstartview=XYZ, plainpages=false}

\makeatletter

\providecommand{\tabularnewline}{\\}

\let\oldforeign@language\foreign@language
\DeclareRobustCommand{\foreign@language}[1]{%
	\lowercase{\oldforeign@language{#1}}}

\makeatother

\allowdisplaybreaks

\makeatletter
\newcommand{\thickhline}{%
	\noalign {\ifnum 0=`}\fi \hrule height 1.5pt
	\futurelet \reserved@a \@xhline
}
\newcolumntype{"}{@{\hskip\tabcolsep\vrule width 1pt\hskip\tabcolsep}}
\makeatother

\newcolumntype{P}[1]{>{\centering\arraybackslash}p{#1}}
\newcolumntype{M}[1]{>{\centering\arraybackslash}m{#1}}

\begin{document}
	%
	\title{$N-1$ Reliability Makes It Difficult for False Data Injection Attacks to Cause Physical Consequences}

	
	
	%
	\author{\IEEEauthorblockN{Zhigang Chu,
			Jiazi Zhang, 
			Oliver Kosut, and
			Lalitha Sankar\\}
		\IEEEauthorblockA{School of Electrical, Computer and Energy Engineering\\
			Arizona State University}}


	\maketitle
	\pagestyle{plain}
	\begin{abstract}
		This paper demonstrates that false data injection (FDI) attacks are extremely limited in their ability to cause physical consequences on $N-1$ reliable power systems operating with real-time contingency analysis (RTCA) and security constrained economic dispatch (SCED). Prior work has shown that FDI attacks can be designed via an attacker-defender bi-level linear program (ADBLP) to cause physical overflows after re-dispatch using DCOPF. In this paper, it is shown that attacks designed using DCOPF fail to cause overflows on $N-1$ reliable systems because the system response modeled is inaccurate. An ADBLP that accurately models the system response is proposed to find the worst-case physical consequences, thereby modeling a strong attacker with system level knowledge. Simulation results on the synthetic Texas system with 2000 buses show that even with the new enhanced attacks, for systems operated conservatively due to $N-1$ constraints, the designed attacks only lead to post-contingency overflows. Moreover, the attacker must control a large portion of measurements and physically create a contingency in the system to cause consequences. Therefore, it is conceivable but requires an extremely sophisticated attacker to cause physical consequences on $N-1$ reliable power systems operated with RTCA and SCED.
	\end{abstract}
	

	%
	\IEEEpeerreviewmaketitle
	\global\long\def\figurename{Fig.}
	\global\long\def\tablename{TABLE}
	
	\vspace{-0.8cm}
	\section{Introduction}
    The efficiency and intelligence of modern electric power systems are increasing rapidly with integration of real-time monitoring, sensing, communication and data processing. This integration is accomplished via a cyber layer consisting of the supervisory control and data acquisition (SCADA) system in conjunction with the energy management system (EMS). SCADA monitors the physical system, collects measurements, and sends them to the control center. In the EMS, state estimation (SE) estimates the voltage magnitudes and angles from measurements. This estimate along with the subsequent data processing, optimization and communication, specifically real-time contingency analysis (RTCA) \cite{Mittal2011} and security constrained economic dispatch (SCED) \cite{SCED2006}, allow for real-time control of the power systems. 
	
	However, the integration of the cyber layer also increases the threat of cyber-attacks on power systems that could lead to severe physical consequences, as illustrated by the recent cyber-attack in Ukraine (see \cite{UkraineAttack}). Therefore, it is crucial to develop techniques to detect and thwart potential attacks, which requires evaluating system vulnerability to credible attacks. Assessing consequences of possible attacks is extremely instructive for system operators, and is important for secure power system operations. 
	
	\textit{Related work:} This paper focuses on unobservable false data injection (FDI) attacks, wherein a malicious attacker replaces a subset of SCADA measurements (power flows and injections) with counterfeits. A wealth of research effort has been undertaken on FDI attacks, showing that they can be designed to target system states \cite{Liu2009,Kosut2011,Hug2012}, system topology \cite{Jzhang2016, Liu2017}, and energy markets \cite{Moslemi2018}. They can bypass the bad data detector (BDD) embedded within SE, and change the load data used for re-dispatch, which in turn cause physical and/or economic consequences. Many existing work evaluating the worst-case attack consequences involve solving attacker-defender bi-level linear programs (ADBLPs), wherein the first level models the attacker's objective and limitations (\textit{e.g.}, number of measurements to change), while the second level models the system response under attack via DC optimal power flow (OPF). Examples include attacks that cause line overflows \cite{Liang2015}, locational marginal price (LMP) changes \cite{Jia2014}, operating cost increases \cite{Yuan11} and sequential outages \cite{Che2019}. The authors of \cite{Zhang2018} analyzes the physical consequences when the attacker only has limited information, and \cite{Chung2019} and \cite{Xiang2017} focus on cyber-physical coordinated attacks. The authors of \cite{Kang2018} propose an ADBLP to find FDI attacks that add or drop contingency pairs with minimum attack effort, and analyze the economic effect of such attacks on LMPs. Rahman \textit{et al.} \cite{Rahman2019} demonstrate several case studies to showcase the impact of FDI attacks on contingency analysis, but their approach is not optimization-based, which means that it does not consider worst-case scenarios. Both \cite{Kang2018} and \cite{Rahman2019} consider simplified SCED as system response, but the only addition of their SCED to DCOPF is the contingency case line power flow constraints modeled using DC line outage distribution factors (LODFs), while other SCED constraints such as reserve and ramp rate constraints are not considered. 

	Despite this prior research, there remains a need to evaluate physical consequences of FDI attacks that take into account \textit{detailed} models for the system response, including RTCA and SCED. To understand the worst case consequences, we evaluate the vulnerability of $N-1$ reliable power systems, by modeling a powerful attacker that has system level knowledge and capabilities. In particular, we focus on unobservable FDI attacks that aim to maximize the power flow on a target line after re-dispatch \cite{Liang2015}. The authors of \cite{Liang2015} design such attacks by solving an ADBLP modeling DCOPF as the system response and demonstrate that they can cause physical overflows. However, we found in our experiments that these attacks fail to cause overflows on systems operating with RTCA and SCED. This observation leads to another question: can attacks designed with complete knowledge of operations lead to more consequences? Note that answering this question inherently focuses on very strong attackers, as in general there is no universally adopted formulation of SCED, and we assume the attacker knows the SCED formulation for the particular system that it is attacking. Our goal of modeling such strong attackers is to understand if the grid is resilient to such worst-case attacks. The resulting attacks are tested on the synthetic Texas system \cite{TexasSystem} with 2000 buses to demonstrate the difficulty of causing physical consequences. Our results show that $N-1$ reliability achieved by RTCA and SCED leads to more conservative operation, making it hard for FDI attacks to cause any pre-contingency overflows, even for the above-mentioned strong attacker. The attacks may still cause post-contingency overflows, but this requires the attacker to perform a cyber-physical coordinated attack by physically creating a contingency. Furthermore, as we show later, these sophisticated attacks also require the attacker to control measurements in a large portion of the system, which is again difficult to achieve in practice.
	
	To summarize, the key contributions of this paper are as follows:  
	
	1. We showcase that attacks designed without considering EMS operations including RTCA and SCED do not cause the physical consequences intended by the attacker.
		
	2. Given this observation, we propose an ADBLP modeling SCED as the system response, assuming an extremely strong attacker who has perfect knowledge of EMS operations including RTCA and SCED.
		
	3. We provide simulation results on the synthetic Texas system with 2000 buses. We find that the resulting attacks can only cause post-contingency overflows. 
		
	4. We highlight that even the aforementioned powerful attacker must control a large number of measurements and physically create a contingency to cause overflows. 

	The remainder of this paper is organized as follows. Sec. \ref{sec: SE and models} describes the power system measurement model and unobservable attack model. Sec. \ref{sec:OPFAtk} demonstrates that attacks designed with DCOPF are extremely limited in their ability to have expected consequences if the system re-dispatch using RTCA and SCED. Sec. \ref{sec:worstAtk} details the knowledge and capabilities of the worst-case attacker, and introduces an ADBLP modeling SCED as system response to find worst-case attacks. Sec. \ref{sec:Simulation} illustrates the simulation results on the synthetic Texas system. Concluding remarks and future work are presented in Sec. \ref{sec:conclusion}.
	\vspace{-0.7cm}
	
	\section{\label{sec: SE and models}System and Attack Model}
	\subsection{EMS Operation\label{sec:EMS_OP}}
\begin{figure}[h]
		\centering{}\includegraphics[trim=0 0.2cm 0 0.2cm, scale=0.45]{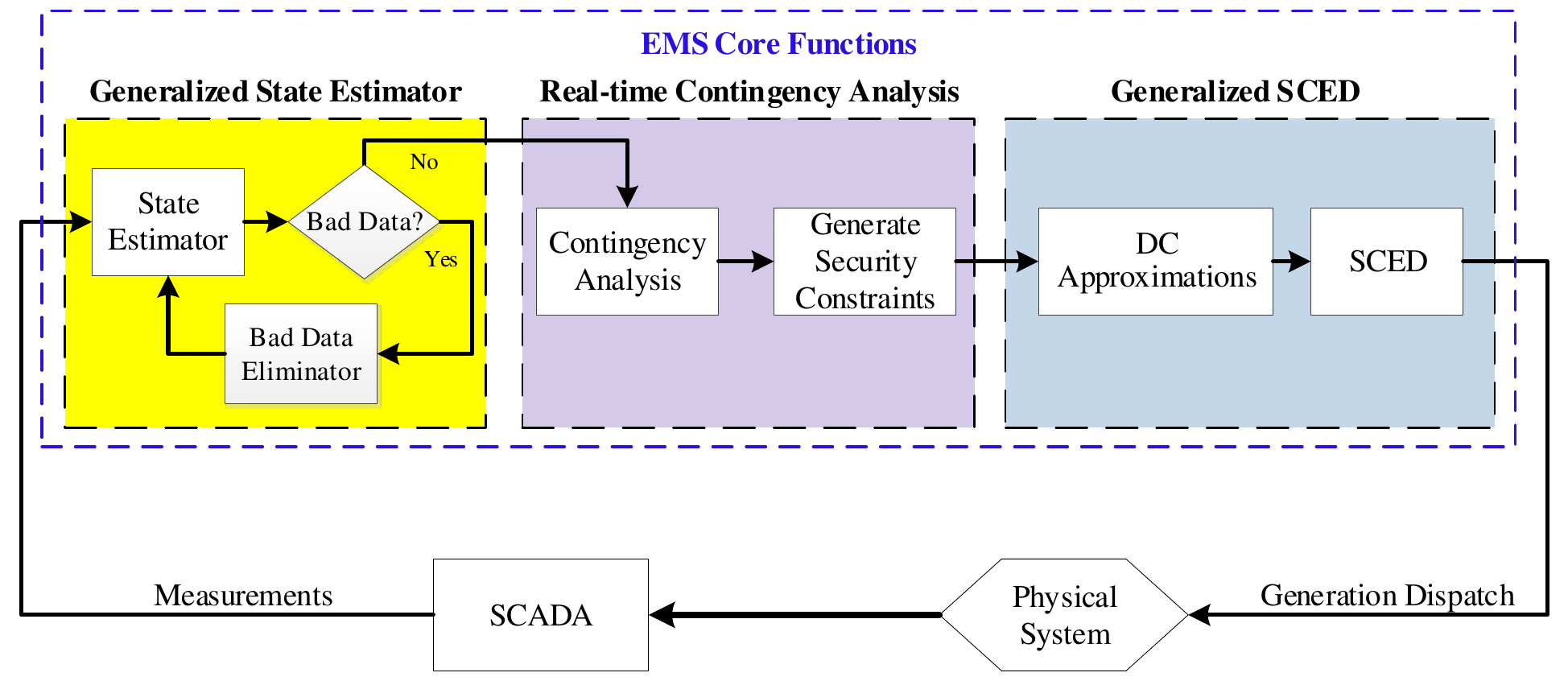}\protect\protect\caption{EMS operation with SE, RTCA, and SCED. \label{fig:EMS_op}}
		\vspace{-0.5cm}
	\end{figure}

	In this paper, we consider an EMS with three core functions operating in the order of SE, RTCA, and SCED. The EMS operating structure is illustrated in Fig. \ref{fig:EMS_op}. Power system measurement data collected by SCADA are sent to SE, which estimates the complex voltages after eliminating noise and bad measurements. Given the generator set points, load values are estimated based on SE results. Modern power systems typically require $N-1$ reliability, \textit{i.e.,} the system must operate with no violations if a contingency occurs (one of the system components, generators or branches, is out of service). RTCA simulates one power flow under each contingency $k$. We say a branch has a \textit{warning} if its power flow is above a threshold $\tau$ but less than its limit, while a branch has a \textit{violation} if its power flow exceeds its limit. Both ``warning'' and ``violation'' branches are denoted \textit{critical branches}. For post-contingency critical branches, their corresponding contingencies are called \textit{critical contingencies}. Note that in base case, the branch limits are the long-term ratings, while in contingency case they are the short-term ratings. RTCA generates one security constraint to be modeled in SCED for each warning and violation. SCED takes all security constraints, along with other common constraints including reserve and ramp rates, to solves an optimization problem to determine the most economic generation dispatch that ensures $N-1$ reliability.
	\vspace{-0.3cm} 
	\subsection{Measurement Model}
	We model the power system with $n_b$ buses, $n_g$ generators, and $n_m$ measurements. 
	The SCADA system measurement model is given by
	\begin{equation}
		z=h(x)+e\label{eq:DCMeasurement}
	\end{equation}
	where $z$ is the $n_{m}\times1$ measurement vector; $x$ is the $2n_b\times1$ vector of bus voltage magnitudes and angles (states); $h(\cdot)$ is the non-linear relationship between measurements and states; $e$ is the $n_{m}\times1$ vector of measurement noise, whose entries are assumed to be jointly distributed as $\mathcal{N}$(0,$R$) where $R=\text{diag}(\sigma_{1}^{2},\sigma_{2}^{2},\ldots,\sigma_{n_{m}}^{2})$.
	\subsection{\label{AttackModel}Unobservable Attack Model}
	An $n_{m}\times 1$ false measurement vector $\bar{z}$ is defined to be \emph{unobservable} to the traditional residual-based BDD if 
	\begin{equation}
		\bar{z}=h(x+c)+e, \label{eq:falseMeas}
	\end{equation} 
	where $c$ is the attack vector \cite{Hug2012}. These false measurements cannot be distinguished from the true measurements if the true states are $x+c$, and hence, cannot be detected by the traditional BDD. Given $c$, an attack subgraph $\mathcal{S}$ can be constructed as in \cite{Hug2012}, such that the non-zero entries of $\bar{z}-z$ are all within $\mathcal{S}$. Launching such an attack requires the attacker to gain control of all measurements within $\mathcal{S}$.
%
%
	By modifying measurements in $\mathcal{S}$, the attacker can arbitrarily spoof the states of center buses (load buses corresponding to non-zero entries of $c$) without detection. The attack causes the system estimated loads to re-distribute between load buses within $\mathcal{S}$, while the total load remain unchanged.


	\section{Consequences of Attacks Designed with DCOPF on $N-1$ Reliable System \label{sec:OPFAtk}}
	In this section, we demonstrate that attacks designed without considering RTCA and SCED (as in many existing literatures) do not cause expected physical consequences on systems operated as outlined in Fig. \ref{fig:EMS_op}. The attacker's capability assumptions and the attack design ADBLP are adopted from \cite{Liang2015}. The purpose of the attacker is to maximize the physical power flow on a target line after re-dispatch, and possibly cause overflow. The attacker is assumed to have knowledge of: (i) the complete network topology (including line parameters and ratings) and load information, and (ii) the cost, capacity, and operational status of all generators in the system. The formulation of this ADBLP is given by
	\begin{subequations}\label{eq:OPFAtk}
		\begin{flalign}
		\hspace{-0.6cm}\underset{c}{\text{maximize}}\: \hspace{0.2cm} & P_{l}-\sigma\left\Vert c\right\Vert _{1}\label{eq:Obj1_MaxPF_OPF}\\
		\text{subject to} \hspace{0.2cm}\; \notag  \\
		& \hspace{-1.1cm}\left\Vert c\right\Vert _{1}\le N_{1}\label{eq:con_resources_OPF}\\
		& \hspace{-1.1cm}-L_{S} P_{D}\le Hc\le L_{S} P_{D}\label{eq:con_loadshift_OPF}\\
		& \hspace{-1.1cm}P_l=\text{PTDF}^l(G_{B}P_{G}^{*}-P_{D}) \label{eq:Physical_PF_OPF}\\
		& \hspace{-1.1cm}\left\{P_{G}^{*}\right\} =\text{arg}\left\{ \underset{P_{G}}{\text{min}}\: C_{G}\left(P_{G}\right)\right\} \label{eq:OBJ_MINCOST_OPF}\\
		&\notag \hspace{-1cm} \text{subject to}\\
		&\hspace{-0.6cm}\begin{array}{lr} 
		\sum_{g=1}^{n_{g}}P_{Gg}=\sum_{i=1}^{n_{b}}P_{Di} & \end{array}\label{eq:con_nodebalance_OPF}\\
		&\begin{array}{lc}
		\hspace{-0.6cm}-P_\text{max}\le \text{PTDF}(G_{B}P_{G}-P_{D}+Hc)\le P_\text{max} \end{array}\label{eq:con_powerflow_OPF}\\
		& \begin{array}{cc}
		\hspace{-0.6cm} P_{G,\text{min}}\le P_{G}\le P_{G,\text{max}} &\end{array}\label{eq:con_GENlimit_OPF}
		\end{flalign}
	\end{subequations}
	where the variables are:
	\begin{description}[leftmargin=1.8cm,style=multiline]
		\item[$c$] attack vector, $n_b\times 1$;
		\item[$P_{l}$] physical power flow on target line $l$;
		\item[$P_{G}$] power output of generators, $n_g\times 1$;
	\end{description}
	and the parameters are:
	\begin{description}[leftmargin=1.8cm,style=multiline]
		\item[$\sigma$] penalty of the $l_1$-norm of attack vector $c$;
		\item[$G_{B}$] generators to buses connectivity matrix, $n_{b}\times n_{g}$;
		\item[$N_{1}$] attack vector $l_{1}$-norm limit;
		\item[$L_{S}$] load shift factor, in percentage;
		\item[$H$] dependency matrix between power injection
		measurements and states, $n_{b}\times n_{b}$;
		\item[$P_{D}$] vector of real loads, $n_{b}\times1$;
		\item[$C_{G}$] generation cost vector, $n_g\times 1$;
		\item[$\textnormal{PTDF}$] power transfer distribution factor matrix;
		\item[$\textnormal{PTDF}^l$] the $l^\text{th}$ row of PTDF matrix;
		\item[$P_{\max}$] vector of base case line limits;
		\item[$P_{G,{\min}}$]  generation lower limits vector, $n_g\times 1$;
		\item[$P_{G,{\max}}$]  generation upper limits vector, $n_g\times 1$.
	\end{description}

	In DCOPF, the voltage magnitudes are all considered to be 1 p.u., and hence, $c$ is an $n_b\times 1$ attack vector on the voltage angles. The objective function \eqref{eq:Obj1_MaxPF_OPF} is to maximize the physical power flow on target line $l$, and the second term penalizes the $l_1$-norm of attack vector $c$, such that if there exists multiple optimal solutions, the one with the smallest $\|c\|_1$ will be selected. Constraint \eqref{eq:con_resources_OPF} limits the attacker's resources. Ideally, this should be characterized by the number of states that can be changed by the attacker, which is the $l_0$-norm of $c$. However, $l_0$-norm is non-convex and intractable, here we use $l_1$-norm as a proxy. \eqref{eq:Physical_PF_OPF} calculates physical power flows from the optimal generation dispatch under attack. \eqref{eq:con_loadshift_OPF} characterize the detectability of the attacks in terms of load shift, because loads that deviating too much from their true values are easily detectable. Note that $Hc$ is a DC approximation of the injection measurement changes caused by the attack, because the AC relationship $h(\cdot)$ is non-convex. \eqref{eq:OBJ_MINCOST_OPF}-\eqref{eq:con_GENlimit_OPF} are DCOPF under attack. It is illustrated in \cite{Liang2015} that the attacks obtained by solving \eqref{eq:OPFAtk} can cause physical overflows if the system re-dispatches using DCOPF.
	
	
	However, modern EMSs typically operate as outlined in Fig. \ref{fig:EMS_op}. Thus, the attacker cannot accurately predict the system response by solving \eqref{eq:OPFAtk}, and the re-dispatch after attack may not cause expected consequences. We have found in our experiments that attacks designed with DCOPF cannot cause any overflows on the synthetic Texas system operating with RTCA and SCED \textit{even in the peak load scenario}. To illustrate this, consider the following example from our experiments.
	
\begin{figure}[h]
		\centering{}\includegraphics[trim=0 0.2cm 0 0.2cm, scale=0.5]{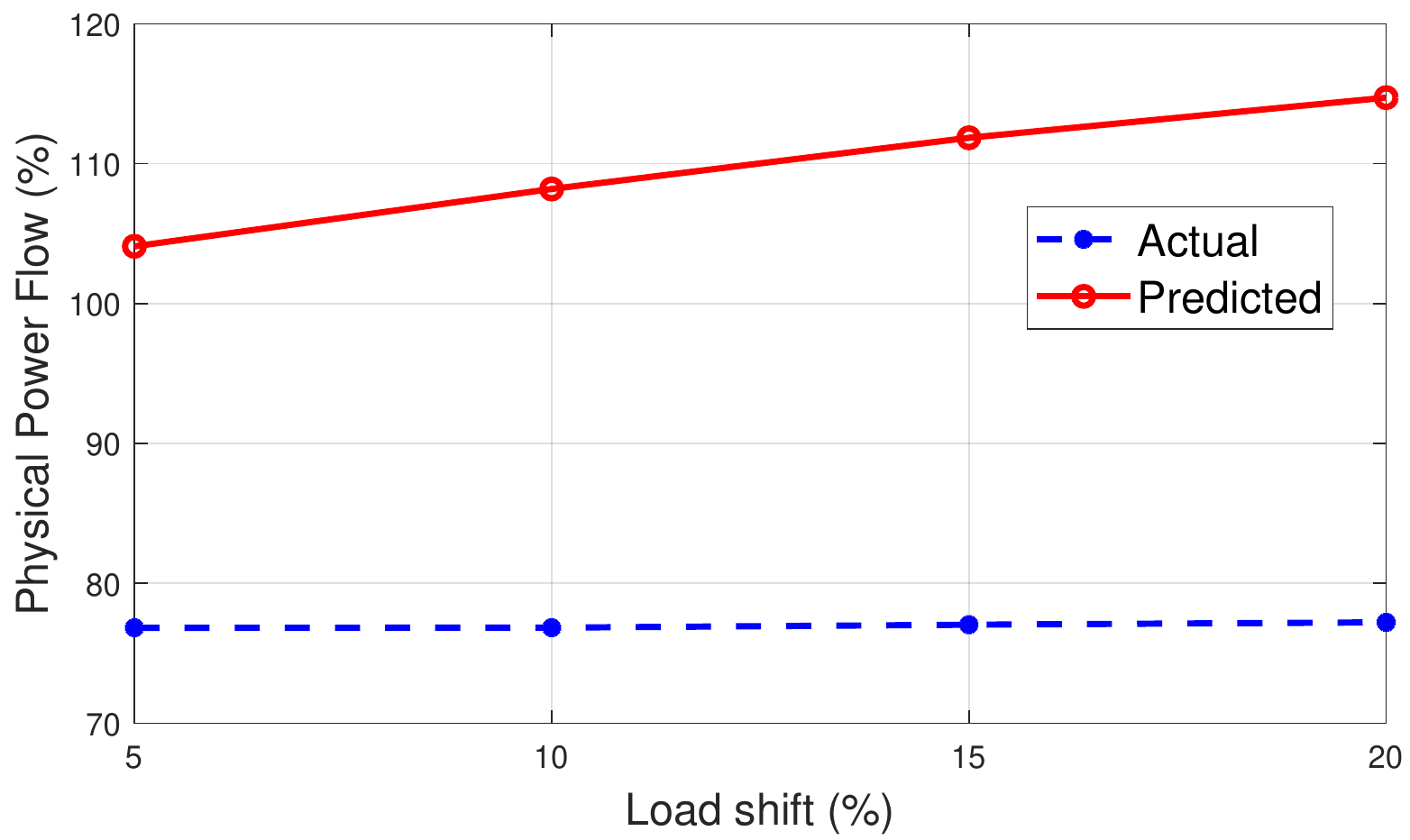}\protect\protect\caption{Consequence of attacks designed with DCOPF on $N-1$ reliable synthetic Texas system, $N_1=2$. \label{fig:OAresult}}
				\vspace{-0.3cm}
	\end{figure}

	The attacker continuously monitors the system operating status, and at the peak load hour, it observes that the most critical branch is transformer ``tx-3083-3082'' with a power flow of 76.72\%. It selects this branch as the target and uses \eqref{eq:OPFAtk} (that is, modeling the system response via DCOPF) to obtain the attack vector $c$ as well as the predicted physical power flow. It finds that the predicted flow exceeds the rating. Hence, it creates false measurements $\bar{z}=h(\hat{x}+c)$ to launch an attack. The system estimates loads from $\bar{z}$, and performs RTCA and SCED to find the optimal generation dispatch (details of RTCA and SCED are given in Sec. \ref{sec:ADBLP}). Applying the new dispatch on the real loads yields the actual physical power flows. Fig. \ref{fig:OAresult} illustrates a comparison between the attacker's predicted physical power flows and the actual flows on this target branch as a function of load shifts $L_S$, with $N_1=2$.
	\vspace{-0.2cm}
	
 	From this figure, we can see that the attacker predicted power flows exceed the rating of the branch for every load shift, but the actual flows are not affected. This is because in the pre-attack DCOPF solution, the target branch is congested. The attack redistributes the loads in the system, making it appear that the flow on this branch is reduced. The higher the load shift, the more the reduction on the flow. Thus, DCOPF will re-dispatch the generations to increase the flow on this branch, making it congested again. This will overload the branch in the physical system, since the real loads are not changed. However, SCED models more constraints than DCOPF does, and this branch is congested in neither base case nor contingency cases. The load redistribution caused by the attack does not affect any binding constraints in SCED, and hence, has no effect on the re-dispatch. We have experimented on the 5 branches with highest base case flows, and observed similar consequences. 
 	


	\vspace{-0.3cm}
	\section{\label{sec:worstAtk}Worst-case Attacks}
	\subsection{Attacker Assumptions\label{sec:ATK_assumption}}
	The observations illustrated in the previous section lead to the following new question: if the attacker knows the system operation details, can it cause physical consequences through FDI attacks? To this end, we model the worst-case attacker who has knowledge of system EMS operations. In other words, the attacker is able to perform the same RTCA and SCED as the system does, and hence, can design attacks that maximize the consequences. This is a very strong assumption, because in addition to having access to the database of the control center, now the attacker further knows the algorithms and assumptions used by the system. While this assumption may be impractical, we aim to understand whether $N-1$ reliable system is resilient against such strong adversaries through this worst-case approach.
	
	In order to accurately predict the system response under attack, the attacker needs to know all the constraints modeled in SCED. This requires the attacker to gain knowledge of the power flow algorithm used in RTCA to get the same post-contingency flows on all branches, as well as the threshold $\tau$ as described in Sec. \ref{sec: SE and models}, to determine the security constraints to be included in SCED. In addition to these security constraints, the attacker must know the detailed modeling of other SCED constraints, as different systems may have different SCED implementations. We assume the attacker has full knowledge of RTCA and SCED implementations, in particular:
	\begin{enumerate}
		\item Contingency ratings of the branches;
		\item Loss handling method;
		\item Ramp rates and reserve costs of all generators;
		\item Reserve policy and requirements;
		\item Criteria to determine which base case line limits are to be modeled. This can be the same threshold as $\tau$ in post-contingency case, but can also be different; 
		\item Branch flow calculation method in both base case and contingency case;
		\item Load shedding policy and costs.
	\end{enumerate}
	Although it is not entirely impossible to have this level of knowledges \cite{Liang2017}, since such complex attacks often involve sophisticated (even nation-state) attackers that can exploit or have access to insider knowledge \cite{UkraineAttack,StuxnetIran}, it is still extremely hard to have such a strong attacker in practice. However, this is the worst-case assumptions from the optimization perspective, because the attacker can most accurately predict the system response by modeling exactly same SCED. Modeling less constraints or relaxing any of those constraints will increase the feasible region of SCED, and hence, exaggerate the attacker predicted consequences. Understanding the vulnerability of power system to such worst-case attacks can serve as an upper bound on risks to system operations.
	
	\vspace{-0.3cm}
	\subsection{ADBLP to Find Worst-case Attacks\label{sec:ADBLP}}
	The worst-case line overflow attacks can be found using an ADBLP similar to \eqref{eq:OPFAtk}. The first level models the attacker's objective and limitations, while the second level models the system response via SCED. The security constraints are generated by an RTCA that simulates branch contingencies, excluding radial branches. Contingency $k$ indicates that branch $k$ is out of service. The attacker can choose critical branches in either base case or contingency case as target branch. Without loss of generality, we assume the flow on $l$ is positive; if it is not the case, its absolute value can be maximized.
	
	The ADBLP takes the following form:
	\begin{subequations} \label{eq:ADBLP}
	\begin{flalign}
		\hspace{-0.3cm}\underset{c}{\text{maximize}}\: \hspace{0.2cm} & P_{l}-\sigma\left\Vert c\right\Vert _{1} \hspace{0.3cm} \text{or} \hspace{0.3cm}P_{l,k_t} - \sigma\|c\|_1 \label{eq:Obj1_MaxPF}\\
		\notag \text{subject to}\hspace{0.2cm}\;\\
		& \hspace{-1cm}\|c\|_1 \le N_1 \label{eq:con_L1_norm}\\
		& \hspace{-1cm}-L_{S} P_{D}\le Hc\le L_{S} P_{D}\label{eq:con_loadshift}\\
		& \hspace{-1cm}P_{l}=\text{PTDF}^l(G_{B}P_{G}^{*}-P_{D}) \label{eq:Physical_PF_base}\\
		& \hspace{-1cm}P_{l,k_t}=\text{OTDF}_{k_t}^l(G_{B}P_{G}^{*}-P_{D}) \label{eq:Physical_PF}\\
		& \hspace{-1cm}\left\{P_{G}^{*}\right\} =\text{arg}\left\{ \underset{P_{G},R_G,P,P_k}{\text{min}}\: C_{G}\left(P_{G}\right)+C_RR_G\right\} \label{eq:OBJ_MINCOST}\\	
		&\notag \hspace{-0.9cm} \text{subject to}\\	
		& \hspace{-0.3cm}\begin{array}{lr} 
		\sum_{g=1}^{n_{g}}P_{Gg}=\sum_{i=1}^{n_{b}}P_{Di}\end{array}\label{eq:con_nodebalance}\\
		& \hspace{-0.2cm}\bar{P}=P_0+\text{PTDF}(G_B(P_G-P_{G0})+Hc)\label{eq:con_basePF}\\
		& \hspace{-0.2cm}\bar{P}_k=P_{k0}+\text{OTDF}_k(G_B(P_G-P_{G0})+Hc)\label{eq:con_ctgcPF}\\
		& \notag \hspace{0.6cm} + \text{LODF}_k\cdot\text{PTDF}^k\cdot Hc, \forall k\\
		& \hspace{-0.2cm}-P_\text{max}\le \bar{P}\le P_\text{max} \label{eq:con_basePmax}\\
		& \hspace{-0.2cm}-P_{k,\text{max}}\le \bar{P}_k\le P_{k,\text{max}}, \forall k \label{eq:con_ctgcPmax}\\
		& \hspace{-0.2cm} P_G \ge \text{max}\{P_{G0}-M_GT_h, P_{G,\text{min}}\}\label{eq:con_PGmin}\\
		& \hspace{-0.2cm} P_G \le \text{max}\{P_{G0}+M_GT_h, P_{G,\text{max}}\}\label{eq:con_PGmax}\\
		& \hspace{-0.2cm} 0 \le R_G \le M_GT_r\label{eq:con_reserve}\\
		& \hspace{-0.2cm} P_G+R_G\le P_{G,{\text{max}}}\label{eq:con_genlimit}\\
		& \hspace{-0.2cm} \begin{array}{lr} 
		\sum_{g=1}^{n_{g}}R_{Gg}\ge P_{Gg}+R_{Gg}\end{array}, \forall g\label{eq:con_genctgc}
	\end{flalign}
	\end{subequations}
	In addition to the variables and parameters introduced in \eqref{eq:OPFAtk}, the new notations are listed as follows.
	\begin{description}[leftmargin=1.8cm,style=multiline]
		\item[$\bar{P},\bar{P}_k$] vectors of monitored line cyber power flows in base case and under contingency $k$, respectively;
		\item[$P_{l,k_t}$] physical power flow on target line $l$ under target contingency $k_t$;
		\item[$R_G$] spinning reserve of the generators, $n_g\times 1$;
		\item[$\textnormal{OTDF}_k$] outage transfer distribution factor matrix under contingency $k$;
		\item[$\textnormal{OTDF}^l_k$] $l^{th}$ row of $\textnormal{OTDF}_k$;
		\item[$C_{R}$] reserve cost vector, $n_g\times 1$;
		\item[$P_0,P_{k0}$] vectors of pre-SCED monitored line power flows in base case and under contingency $k$, respectively;
		\item[$P_{G0}$] pre-SCED generator outputs, $n_g\times 1$;
		\item[$\textnormal{PTDF}^k$] $k^{th}$ row of $\textnormal{PTDF}$;
		\item[$\textnormal{LODF}_k$] line outage distribution factors of monitored lines under contingency $k$;
		\item[$P_{k,\max}$] vector of line limits under contingency $k$;
		\item[$M_G$] ramp rates of all generators, $n_g\times 1$;
		\item[$T_h$] look-ahead time for one period SCED;
		\item[$T_r$] time for spinning reserve requirement.
	\end{description}
	
	The attacker's limitations \eqref{eq:con_L1_norm}-\eqref{eq:con_loadshift} are the same as those in \eqref{eq:OPFAtk}. \eqref{eq:Physical_PF_base} and \eqref{eq:Physical_PF} are the physical power flows on line $l$ under base case and under target contingency $k_t$, respectively. The second level SCED \eqref{eq:OBJ_MINCOST}-\eqref{eq:con_genctgc} models the system response to the attack. The SCED is a linearized approximation of ACOPF, and we model it in a ``hot start'' fashion to reduce the AC-DC discrepancy. The objective of the operator \eqref{eq:OBJ_MINCOST} is to minimize the total cost, consisting of generation cost and reserve cost; constraint \eqref{eq:con_nodebalance} is the power balance equation; \eqref{eq:con_basePF} is the cyber power flow of the base case monitored lines. Here $P_0$ is the vector of base case pre-SCED branch flows obtained from RTCA, and is non-linearly related to the pre-SCED generation $P_{G0}$. Therefore, only the change in base case branch flows, $\bar{P}-P_0$, are linearly related to the generation change $P_g-P_{G0}$, and the AC-DC discrepancy is less than that if this constraint is modeled as \eqref{eq:con_powerflow_OPF}. Note that this constraint is only modeled for critical lines whose pre-SCED power flow is greater than the threshold $\tau$, \textit{i.e., } $|P_0/P_{\max}|\ge \tau$. This is under the assumption that the line flows will not change dramatically after the SCED re-dispatch, due to the ramping constraints of the generators. Similarly, \eqref{eq:con_ctgcPF} is the cyber power flows on monitored lines under each contingency $k$, where $|P_{k0}/P_{k,\max}|\ge \tau$. Here we assume the base case and contingency case monitoring thresholds are the same. In the right hand side of \eqref{eq:con_ctgcPF}, the first term is the pre-SCED post-contingency flows; the second term is the change of the flows as a result of re-dispatch and false loads; the third term quantifies the amount of power on the monitored lines resulting from the effect of false loads on the contingency line $k$, which is not considered in $P_{k0}$. Constraints \eqref{eq:con_basePmax} and \eqref{eq:con_ctgcPmax} are the line limits in base case and contingency case, respectively. The active power limits in both base case and contingency cases, $P_{\max}$ and $P_{k,{\max}}$, are approximated from the MVA ratings and reactive flows on the branches by
	\begin{flalign}
		P_{\max} = \sqrt{S_{\max}^2 - [\max(Q_{\text{from}}, Q_{\text{to}})]^2}\label{eq:Pmax}\\
		P_{k,\max} = \sqrt{S_{k,\max}^2 - [\max(Q_{k,\text{from}}, Q_{k,\text{to}})]^2}\label{eq:Pkmax}
	\end{flalign}
	where $S_{\max}$ and $S_{k,\max}$ are branch long-term and short-term ratings, respectively; $Q_{\text{from}}$ and $Q_{\text{to}}$ are the base case reactive branch flows at the "from" end and "to" end, respectively; $Q_{k, \text{from}}$ and $Q_{k, \text{to}}$ are those flows in contingency cases. This is an additional approach to reduce the AC-DC discrepancy.
	Constraints \eqref{eq:con_PGmin} and \eqref{eq:con_PGmax} are the ramp rate limits; \eqref{eq:con_reserve} is the reserve limit; \eqref{eq:con_genlimit} is the generation limit. Though the RTCA does not simulate generator contingencies, in SCED it is required that when a generator is out, the reserves of all other generators are sufficient to cover the output of the lost generator. This system reserve requirement is captured in \eqref{eq:con_genctgc}. 
	
	
	With knowledge of system RTCA and SCED, the attacker can wisely select the target branch, so that the constraints associated with this branch is binding or nearly binding in the pre-attack SCED solution. Therefore, the false loads can mislead the SCED to re-dispatch the generation to increase the flow on the target branch, and possibly cause overflow. Solving the ADBLP \eqref{eq:ADBLP} provides the attack vector $c$ and resulting physical power flows, which allow for evaluating the vulnerability.
	
	\vspace{-0.3cm}
	\subsection{Attack Implementation\label{sec:atk_strategy}}
\begin{figure}[h]
		\centering{}\includegraphics[trim=0 0.2cm 0 0.2cm, scale=0.6]{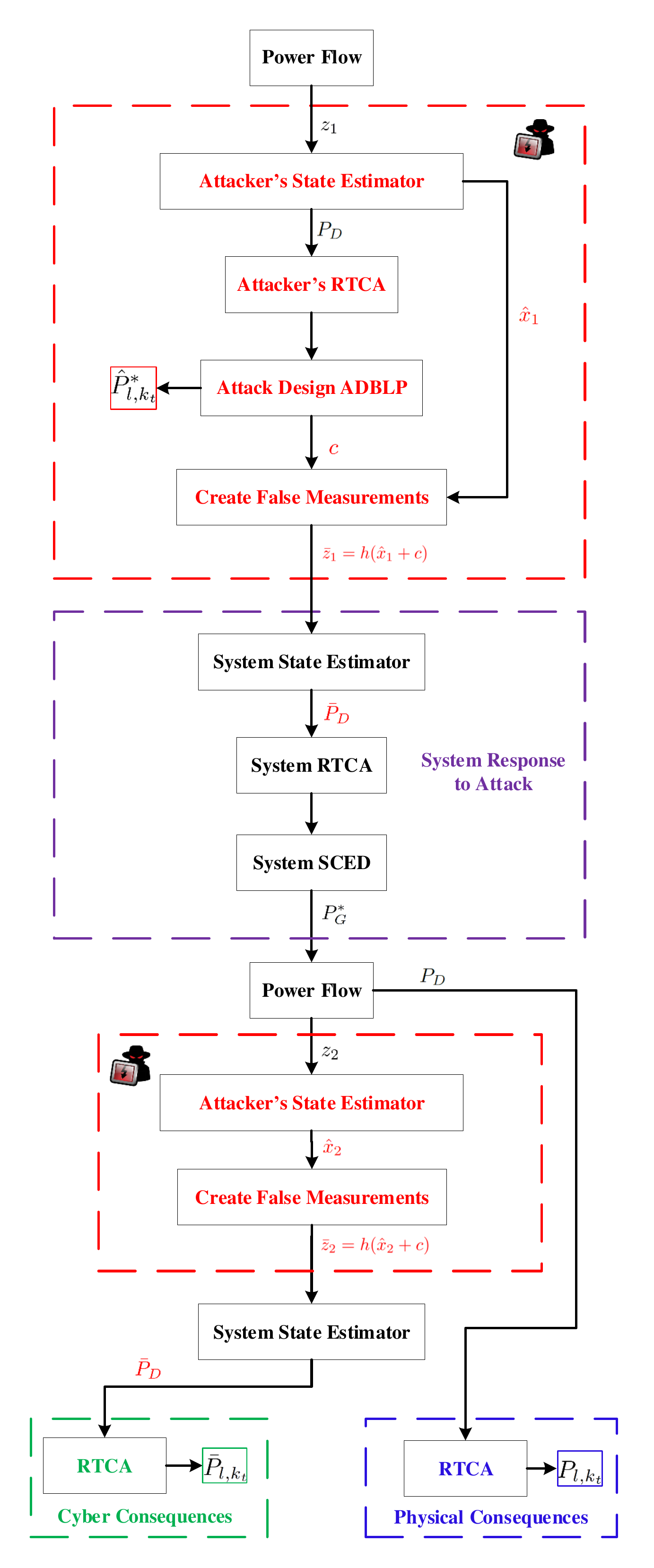}\protect\protect\caption{Attack implementation and system vulnerability assessment approach. \label{fig:atk_strategy}}
				\vspace{-0.6cm}
	\end{figure}

	Fig. \ref{fig:atk_strategy} illustrates the implementation of the attack and the vulnerability assessment approach. We assume the attacker aims to cause post-contingency overflows, and the real loads remain unchanged during the attack period. The physical system behavior and the SCADA measurement collection are simulated by solving an AC power flow. The true measurements $z_1$ from the power flow solution are acquired by the attacker to estimate the states (denoted $\hat{x}_1$). It then performs RTCA to achieve the security constraints and solves the attack design ADBLP to find the attack vector $c$. Recall that the second level of the ADBLP is a SCED in response to the attack, and by solving it the attacker obtains the predicted maximal physical power flow on the target branch, which is the optimal objective ${P}_{l,k_t}^{*}$. To implement the designed attack, the attacker then constructs false measurements $\bar{z}_1=h(\hat{x}_1+c)$ and injects $\bar{z}_1$ to the system SE instead of the true measurements $z_1$. Again, only the measurements in the attack subgraph $\mathcal{S}$ are changed. Since the generator outputs are known to the system, the false measurements will cause the SE to estimate a set of false loads. RTCA and SCED are then performed by the system to determine the new optimal generation dispatch $P_G^*$ in response to the false loads. Once the generators re-dispatch, the true measurements changed to $z_2$. To sustain the attack on the system, the attacker again acquires $z_2$, and estimates the new states $\hat{x}_2$. It then sends $\bar{z}_2=h(\hat{x}_2+c)$ to the system SE to estimate new false loads. The system operator again runs RTCA with the new false loads and observes the cyber power flow $\bar{P}_{l,k_t}$. However, the new dispatch applied on the physical system, will maximize the physical power flow on target line $l$ under target contingency $k_t$, and possibly cause overflow. The true physical power flow, $P_{l,k_t}$, is obtained by running RTCA with the new dispatch on real loads. 
	
	
	\section{\label{sec:Simulation}Simulation Results and Discussions}
	In this section, we present physical consequences through simulations of the attacks designed using ADBLP \eqref{eq:ADBLP}. We use the synthetic Texas system with 2000 buses, 3210 branches, and 432 generators \cite{TexasSystem}. The base case power flow and RTCA are performed using OpenPA \cite{OpenPA}, a Java-based EMS simulation platform that we developed in collaboration with our industry partners IncSys \cite{IncSys} and PowerData \cite{PowerData}. Without attack, the system is operating at steady-state, which means that SCED does not change the generation dispatch between each EMS loop. In the base case power flow solution, the total losses among the system is 2\% of the net load. We assume the SCED handles losses by uniformly increasing all loads by this percentage. RTCA simulates contingencies of all branches whose end bus voltages are both at least 100 kV, except radial branches. The short-term branch limit is assumed to be 115\% of the long-term limit, \textit{i.e.,} $S_{k,\max}=115\%\times S_{\max}$; SCED look ahead time $T_h=15$ minutes; spinning reserve time $T_r=10$ minutes. The ADBLP is solved using a Modified Benders' decomposition (MBD) algorithm that we introduced in \cite{Chu2017}, which can efficiently solve large-scale ADBLPs. With a warning threshold $\tau=90\%$, RTCA reports no base case critical branches, and 25 post-contingency critical branches before attack. We exhaustively design attacks targeting each of those 25 branches for post-contingency overflows with load shift $L_S=[10\%,20\%]$, and $l_1$-norm constraint $N_1=[0.2,2]$ in steps of 0.2. All simulations are conducted on a 3.4 GHz PC with 32 GB RAM.
	
	\vspace{-0.35cm}
	\subsection{Results on Maximal Physical Power Flows\label{sec:results_PF}}
\begin{figure}[h]
		\centering{}\includegraphics[trim=0 0.3cm 0 0.3cm, scale=0.48]{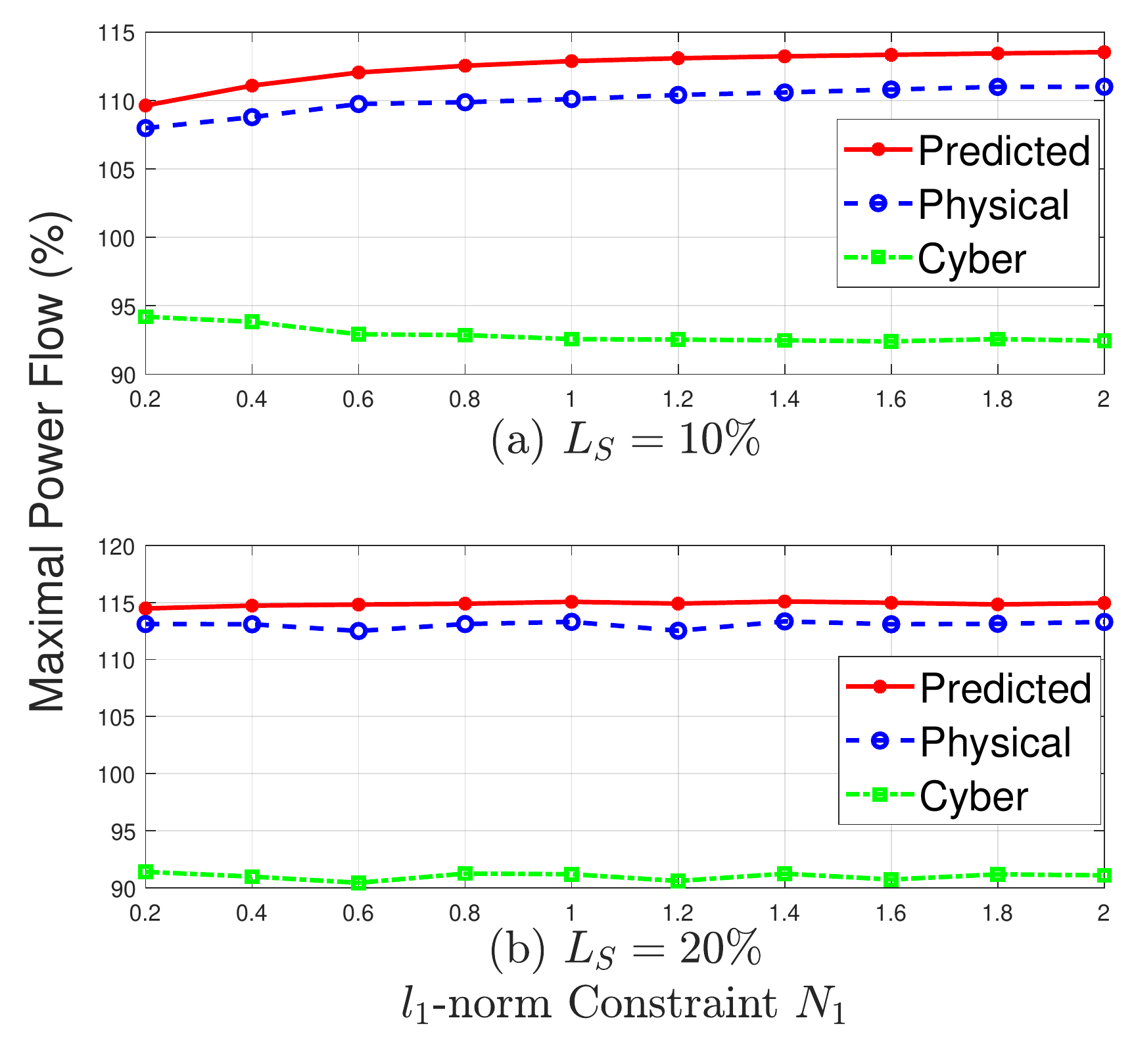}\protect\protect\caption{Comparison of attacker predicted, physical, and cyber power flows on line `ln-2025-2055' under contingency `ln-2054-5236', (a) $L_S=10\%$; (b) $L_S=20\%$ . \label{fig:max_PF}}
	\end{figure}

	Fig. \ref{fig:max_PF} compares physical power flow ${P}_{l,k_t}^{*}$ predicted by the attacker, the true power flow $P_{l,k_t}$ in the physical system, as well as the power flow (cyber) seen by the system operator $\bar{P}_{l,k_t}$, as a function of the $l_1$-norm constraint $N_1$. These power flows are plotted as percentage values relative to the active power limit $P_{l,k,\max}$ calculated using \eqref{eq:Pkmax}. The attacker's goal is to maximize the power flow on line `ln-2025-2055' when line `ln-2054-5236' is out of service. 
	
The results indicate that the attacks cause post-contingency overflows. When the load shift $L_S=10\%$, ${P}_{l,k_t}^{*}$ and $P_{l,k_t}$ increase as $N_1$ increases. When $L_S=20\%$, similar results are observed, but ${P}_{l,k_t}^{*}$ and $P_{l,k_t}$ are not monotonically increasing as $N_1$ increases. This suggests that the MBD algorithm provides sub-optimal solutions, because as $N_1$ increases, the constraints are relaxed, and the optimal solution for a larger $N_1$ should be at least that of a smaller $N_1$. As expected, maximal physical power flow is higher when a larger load shift is allowed.

The true physical power flow $P_{l,k_t}$ is slightly lower than the attacker predicted physical power flow ${P}_{l,k_t}^{*}$. One possible reason for this phenomenon is that the attacker is solving a DC approximation of an AC system, and the reactive power flow may change after attack. This could result in a difference in $P_{l,k,\max}$ before and after attack. Another possible reason is that the false measurements $\bar{z}_1$ injected by the attacker generate a different set of security constraints than those result from true measurements $z$. The attacker uses security constraints generated by pre-attack RTCA to solve the attack design ADBLP, but those constraints used in system SCED are based on the false measurements after attack. As a result, the system SCED solution is different than the attacker predicted re-dispatch. One approach for the attacker to prevent this situation is to run its own RTCA using the false measurements and include any newly appeared security constraints into the attack design ADBLP, until there are no more new security constraints. However, this approach has no convergence guarantee, and could be too time-consuming.

Note that in order for the attacks to actually cause post-contingency violations requires a particular contingency to occur. Thus, the attacker has to physically create the target contingency itself, otherwise it has to wait for the contingency to occur. If the attacker is sufficiently powerful to physically cause contingencies, it may trip multiple lines to shut down the system, and there is no need for cyber-attacks in this situation. As far as we know, the probability of line failure is pretty low in practice. Thus, even though the attacks can cause post-contingency overflows, they can only put the system into an insecure state rather than cause physical damages, because of the difficulty in creating contingencies. 

In the synthetic Texas system, there is no branch whose base case power flow is higher than $\tau$ prior to the attack. Thus, to cause base case overflow, the attacker has to shift a tremendous amount of load that may easily trigger an alarm at the control center. 
We have attempted to design a base case attack targeting top 5 branches with the highest base case power flow in percentage, but no overflow can be found even with $L_S=90\%$ and $N_1=20$. This indicates that RTCA and SCED push the system to operate conservatively, which in turn decreases the vulnerability to line overflow attacks.

\vspace{-0.3cm}
	\subsection{\label{sec:L0result}Results on Attack Resources}
\begin{figure}[h]
		\centering{}\includegraphics[trim=0 0.3cm 0 0.3cm, scale=0.45]{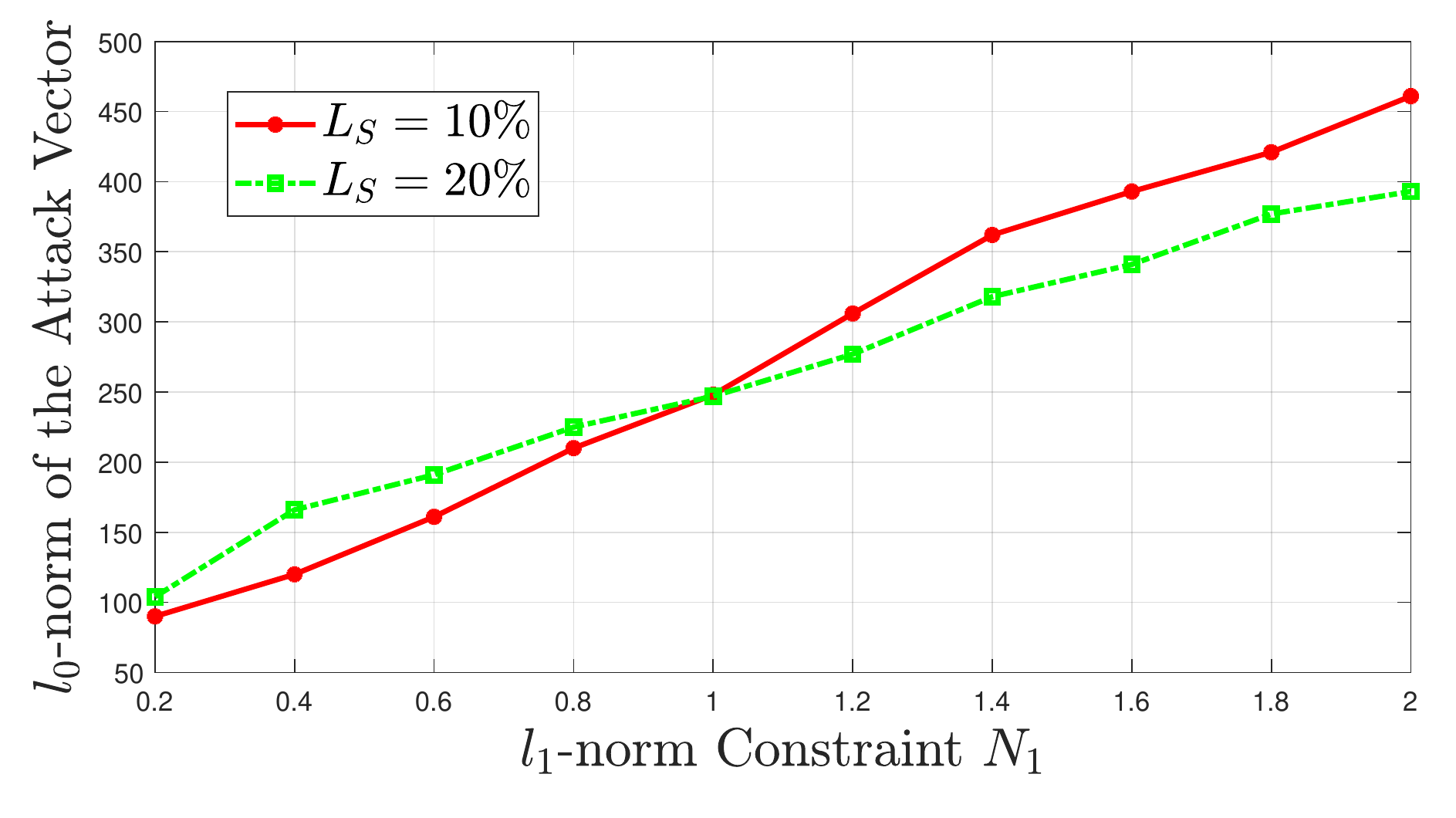}\protect\protect\caption{Comparison of the $l_0$-norm of the attack vector for target line `ln-2025-2055' under contingency `ln-2054-5236'. \label{fig:L0LS}}
		\vspace{-1cm}
	\end{figure}

	Fig. \ref{fig:L0LS} illustrates the relationship between maximal power flow and $l_0$-norm of the attack vector (\emph{i.e.} the number of center buses in the attack) versus the $l_1$-norm constraint $N_1$ for target line `ln-2025-2055' under contingency `ln-2054-5236', with different load shift constraints. As $N_1$ increases, so does the $l_0$-norm of the attack, indicating that $l_1$-norm is a valid proxy for $l_0$-norm for our problem. If a larger load shift is allowed, the maximal power flow on target line increases, but the resulting $l_0$-norm may decrease for the same $N_1$. This indicates a trade-off between load shift and attacker's resources: as the attacker attempts to avoid detection by minimizing load changes, it will require control over a larger portion of the system to launch a comparable attack. These results also indicate that the attacker needs a tremendous amount of resources to launch the attacks. For example, with $L_S=10\%$ and $N_1=1$, the attacker needs to change the state of 250 load buses. The corresponding attack subgraph $\mathcal{S}$ contains more than 800 buses, which is almost half the system. Thus, the attacker must control measurements in almost half of the system to successfully launch this attack, which is extremely hard to achieve.
	\vspace{-0.3cm}
	\subsection{Comparison of Physical and Cyber RTCA results\label{sec:scatter}}
\begin{figure}[h]
		\centering{}\includegraphics[trim=0 0.3cm 0 0.3cm, scale=0.35]{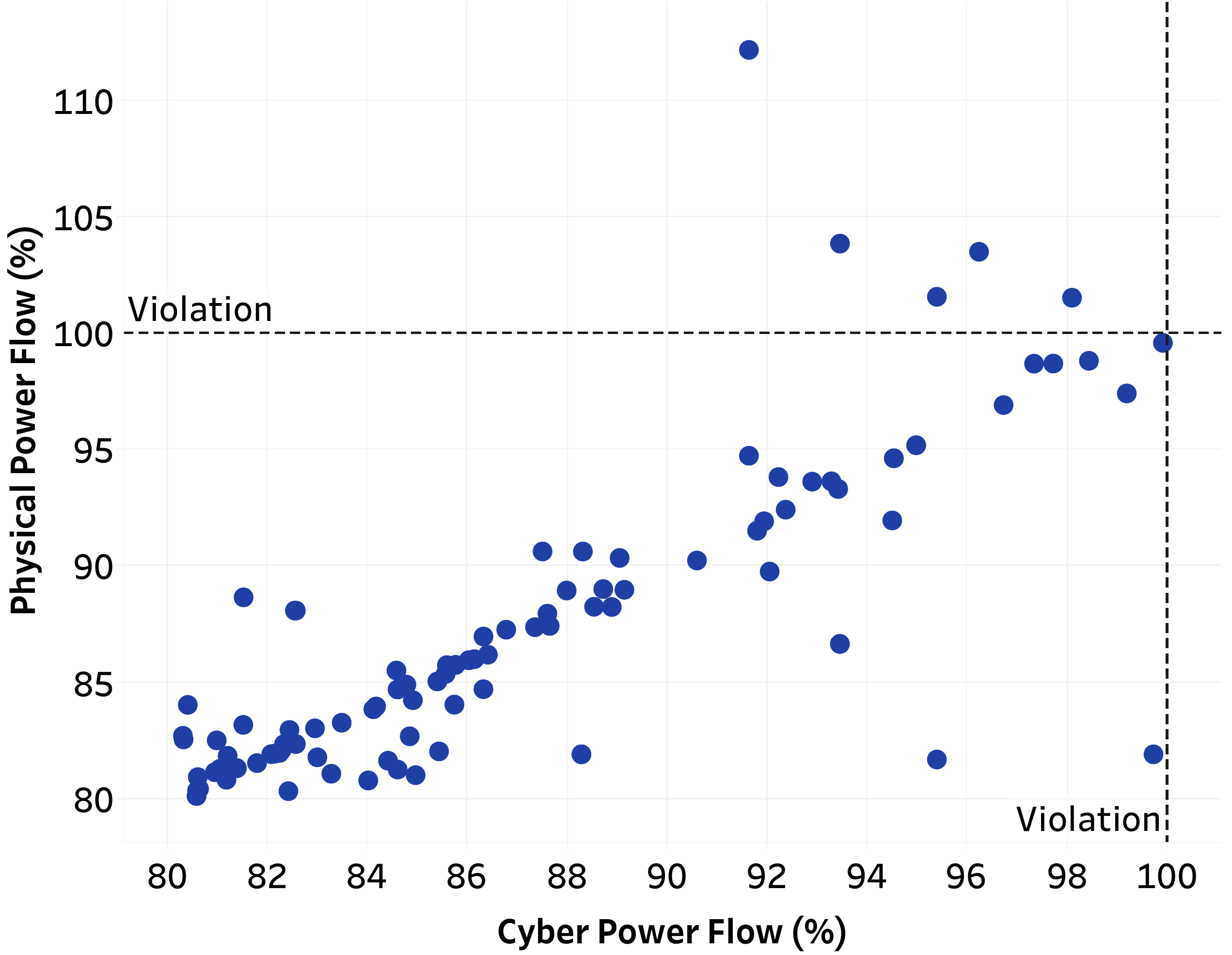}\protect\protect\caption{Physical and cyber RTCA results after re-dispatch. \label{fig:scatter}}
	\vspace{-0.4cm}
	\end{figure}

	Fig. \ref{fig:scatter} compares the physical and cyber RTCA results after the re-dispatch resulting from an attack on target line `ln-2025-2055' under contingency `ln-2054-5236' with load shift $L_S=10\%, N_1=2$. The cyber post-contingency power flows on the x-axis represent what the system operator observes, while the y-axis represents the post-contingency power flows in the physical system. There is no point beyond 100\% of the x-axis, which indicates that the system operator sees no post-contingency violation after the attack. Therefore, the attack successfully spoofed the operator that the system is in a secure state, while in reality, the target line has a 112.2\% post-contingency overflow. In addition, there are four post-contingency violations that are caused by the same attack, even though they are not the attacker's targets, but the overflow percentage are less. This observation indicates that the attack does put the system into an insecure state, and the system is no longer $N-1$ reliable under attack. However, overflows can only occur under contingencies.
	
	\vspace{-0.3cm}
	\vspace{-0.3cm}
	\subsection{Statistical Results on Attack Consequences\label{sec:result_stat}}
	As mentioned at the beginning of Sec. \ref{sec:Simulation}, we exhaustively tested attacks targeting the 25 branches with post-contingency warnings. The designed attacks successfully cause overflows on 8 out of the 25 target branches. Table \ref{tab:result_stat} gives the statistical results on attack consequences of these 8 branches. We derived attacks using $l_1$-norm constraints in the range from $N_1=0.2$ to $N_1=2$. The table shows the resulting ranges in maximal power flow and $l_0$-norm of the attack vector $c$ across this range. The load shift constraint $L_S=10\%$. The prefix `ln' indicates a transmission line and `tx' indicates a transformer. From the maximal power flow range, we can see that some branches are more vulnerable than others, but most of the overflows are within 10\%. Thus, even when the contingencies occur and the target branch becomes overloaded, it still takes time for it to heat up and trip. During this time, the contingencies may be eliminated by the system, and no physical damage can be dealt. Besides, the system operators can identify critical lines and critical contingencies for attack protection purposes. For example, they can artificially
	reduce the line limit to keep the attack from being successful.
	Measurements around vulnerable branches can be encrypted to
	prevent them from being modified. In our ADBLP,
	the load shift constraint characterizes the detectability of
	the attack, indicating that load abnormally detectors can help
	system operators distinguish between natural load changes and
	possible cyber attacks based on load redistribution.
	\begin{table}[h]
		\renewcommand{\arraystretch}{1.3}
		\protect\caption{Statistical Results on Maximal Physical Power Flow and $l_0$-norm of the Attack Vector with $N_1\in [0.2,2]$ \label{tab:result_stat}}
		\centering
		\begin{tabular}{|c|c|c|c|c|c|}
			\thickhline
			\multirow{2}{*}{Target} & \multirow{2}{*}{Contingency} & \multicolumn{2}{|c|}{Max PF (\%)}  & \multicolumn{2}{|c|}{$\|c\|_0$} \tabularnewline
			\cline{3-6}
			& & $N_1$=0.2 & $N_1$=2 & $N_1$=0.2 & $N_1$=2 \tabularnewline
			\hline
			ln-6188-7305 & ln-7058-7095 & 101.92 & 105.08 & 133 & 442 \tabularnewline
			\hline
			ln-6240-6287 & ln-6141-6239 & 102.43 & 106.76 & 137 & 314 \tabularnewline
			\hline
			ln-7233-7251 & tx-6063-6062 & 105.41 & 107.90 & 156 & 485 \tabularnewline
			\hline
			ln-1003-1055 & ln-3046-3078 & 102.80 & 102.94 & 163 & 520 \tabularnewline
			\hline
			ln-2025-2055 & ln-2054-5236 & 107.98 & 111.00 & 90 & 461 \tabularnewline
			\hline
			ln-2070-5237 & ln-2054-5236 & 101.35 & 104.35 & 90 & 461 \tabularnewline
			\hline
			ln-1003-1055 & ln-1004-3133 & 102.43 & 102.56 & 160 & 513 \tabularnewline
			\hline
			ln-7059-7407 & ln-7058-7406 & 100.38 & 102.24 & 154 & 488 \tabularnewline
			\thickhline 
		\end{tabular}
	\end{table}


	
	\vspace{-0.4cm}
	\section{\label{sec:conclusion}Conclusion}
	We have demonstrated that FDI attacks are extremely limited in their ability to cause physical consequences on  power systems operated by EMSs consisting of SE, RTCA, and SCED to ensure $N-1$ reliability. For such systems, we showed that attacks designed with only DCOPF as the system response do not cause expected physical consequences. We then designed attacks by modeling the worst case attacker that can mimic the EMS operations including RTCA and SCED, and tested them on the synthetic Texas system. For this system, we showed that even for the above-mentioned strong attacker, the attacks still cannot cause base case overflows, because the system is pushed to operate conservatively with $N-1$ reliability requirement. The designed attacks can successfully cause post-contingency overflows on target branches, but it requires a specific contingency to occur to deal physical damage to the system. Moreover, the amount of resources required to launch such attacks is tremendous, and the contingencies can be fixed before the overloaded lines trip. Therefore, we argue that it is extremely hard for FDI attacks to cause physical damages to $N-1$ reliable systems. Future work will include designing countermeasures to detect, identify, and mitigate such attacks, to further prevent them from damaging the system.

	\vspace{-0.2cm}
	\section*{Acknowledgment}
	This material is based upon work supported by the National Science Foundation under Grant No. CNS-1449080, and grant S-72 from the Power System Engineering Research Center (PSERC). We would like to thank the following at ASU: Mr. Andrea Pinceti for creating the base case, Mr. Roozbeh Khodadadeh for help with the test platform, and Prof. Kory Hedman and his team for their support with RTCA and SCED. We also thank Dr. Robin Podmore (IncSys) and Mr. Christopher Mosier (Powerdata) for the OpenPA software. 
	\vspace{-0.2cm}

	%
	%

	
	
	%
	%
	%
	
	\bibliographystyle{IEEEtran}
	\bibliography{dis}

\end{document}